# Duration of and time to response in oncology clinical trials from the perspective of the estimand framework

*Hans-Jochen Weber[a], Stephen Corson[b], Jiang Li[c], François Mercier[d], Satrajit Roychoudhury[e], Oliver Martin Sailer[f], Steven Sun[g], Alexander Todd[h], Godwin Yung[i]* on behalf of the Industry Working Group on Estimands in Oncology

## Abstract

Duration of response (DOR) and time to response (TTR) are typically evaluated as secondary endpoints in early-stage clinical studies in oncology when efficacy is assessed by the best overall response (BOR) and presented as the overall response rate (ORR). Despite common use of DOR and TTR in particular in single-arm studies, the definition of these endpoints and the questions they are intended to answer remain unclear. Motivated by the estimand framework, we present relevant scientific questions of interest for DOR and TTR and propose corresponding estimand definitions. We elaborate on how to deal with relevant intercurrent events which should follow the same considerations as implemented for the primary response estimand. A case study in mantle cell lymphoma illustrates the implementation of relevant estimands of DOR and TTR. We close the paper with practical recommendations to implement DOR and TTR in clinical study protocols.

## 1.     Introduction

Early-stage oncology clinical trials often follow a single-arm design to evaluate efficacy of an experimental therapy on the overall response rate (ORR). In such trials, duration of response (DOR) and time to response (TTR) are typically secondary endpoints. Both are presented alongside ORR results in publications and sometimes even in product labels. While study protocols give precise definitions of ORR as the primary estimand, the definitions of DOR, TTR, and the scientific questions they may address, remain unclear.

The purpose of this article is to review DOR and TTR from the perspective of the ICH E9(R1) estimand framework[1]. We start this paper by exploring the scientific questions of interest and considerations surrounding DOR and TTR. In Section 2, we look into clinical considerations of an ORR estimand, which provides the context for the DOR and TTR estimands since DOR and TTR are closely connected with assessing clinical response. In Section 3, we construct typical DOR and TTR estimands and expand to estimand definitions which may be related to scientific questions of interest. In Section 4, the different approaches are illustrated via a case study in mantel cell lymphoma. In section 5, we conclude with practical recommendations regarding the implementation of DOR and TTR in clinical trial protocols and the adoption of alternative DOR and TTR estimands.

---

[a] Novartis Pharma AG, Basel, Switzerland
[b] Phastar, Glasgow, UK
[c] BeiGene, US
[d] Hoffmann-La Roche Ltd, Basel, Switzerland
[e] Pfizer, US
[f] Boehringer Ingelheim Pharma GmbH & Co. KG, Germany
[g] J&J , US
[h] AstraZeneca, UK
[i] Genentech, US

## 2. Overall response rate

The best overall response rate (BOR), also called objective response rate, is commonly used in early-stage oncology clinical trials as the primary endpoint. The overall response rate (ORR) of a treatment group summarizes the BOR of the individual patients whose BOR is CR or PR. Under the assumption that no spontaneous recovery can be expected in the absence of any therapeutic intervention in cancer conditions, any improvement in patient's disease status observed in a clinical trial is attributed to the experimental therapy, thus allowing one to assess efficacy without a control arm. In single-arm trials investigating patients with refractory disease, a response rate endpoint showing a clinically relevant treatment effect size may lead to an accelerated approval[2].

The BOR to a therapy is evaluated by regular response assessments at the patient level. These assessments are based on standardized quantitative and objective criteria as defined in international guidelines specific to the indication of interest like RECIST 1.1 for solid tumors[3]. In the following sections, we focus on the assessment of response as well as on DOR and TTR in solid cancers based on the RECIST 1.1 guidance. Out of all post-baseline response assessments of a patient, the best response category is identified as BOR. Typically, only response assessments which are documented before progressive disease, death and start of new therapies are taken into account: No responses can be observed after death; evidence of progression is the event concluding clinical benefit or having the chance to achieve benefit and any observation after progression is considered not as CR or PR (for immunotherapies please see below). After progression, typically a new anticancer treatment is started. Patients may also start a new anticancer therapy without evidence of progression due to poor tolerability of the initially assigned treatment. RECIST 1.1 is not specific, but the general approach is to consider only responses before the new anticancer therapy is started. This means progression, death and start of new therapies are considered as intercurrent events. Death is addressed by a *while being alive* strategy, progression by *while having no evidence of progression* and start of therapy by *while not having started a new anticancer therapy*. These strategies are referring to the general concept of the "while on treatment" strategy as outlined in the ICH E9 (R1) addendum. Only patients who achieved partial response (PR) or complete response (CR) are considered responders. The first documentation of partial (PR) or complete response (CR) is dated at the onset of response (Figure 1). Confirmation of response by a subsequent response assessment might be required particularly in single-arm studies. Under such a requirement, the onset of response is the date of the first PR or CR that is confirmed subsequently. The summary measure ORR is defined as the proportion of patients who have achieved a CR or PR as their BOR at the population level.

Typically, in early-stage oncology trials, response assessments are performed locally by the investigator and assessments are continued until progression of disease is observed, i.e., when no further treatment benefit can be anticipated. An exception is immunotherapy which may induce effects mimicking progression of disease[4]. In these cases, clinical benefit from continuing immunotherapy can still be expected and meeting criteria of progression for the first time does not necessarily mean failure of the therapy. Thus, response assessments might continue beyond unconfirmed progression in such settings (iRECIST[5]). Unconfirmed progression (iUPD) is addressed as an intercurrent event by the *treatment policy strategy*. BOR as per iRECIST (iBOR) can be observed after unconfirmed progression but not later than confirmation of progression (iCPD).

To avoid the risk of assessment bias in open-label or single-arm trials, response is often assessed by an independent review committee (IRC). The IRC and the investigator may come to different conclusions[6]. Since investigator assessments are used to guide treatment decisions, it may not be possible to follow patients for response if the investigator determined progression. This has also impact on DOR and TTR estimands if based on IRC assessment.

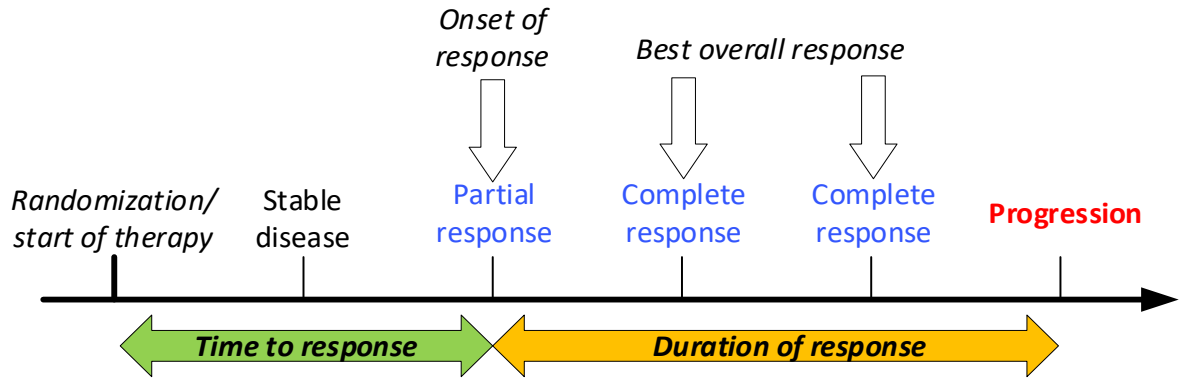

**Figure 1**: Response assessments: Example

In the following, we discuss clinical assumptions which are implied in a typical ORR estimand as based on BOR and which provide relevant context to considerations regarding DOR and TTR:

- **Measurable disease at baseline is a pre-requisite to assess response.** Patients can only be assessed appropriately for response post-baseline if they present with measurable disease at baseline. Response is typically assessed by comparing the post-baseline measurement with the baseline status. Absence of all lesions post baseline can only be considered clinical response in relation to a measurable baseline lesions. Accordingly, measurable disease at baseline is a relevant pre-requisite for TTR and DOR as well.

- **BOR is agnostic of the time of onset.** A patient is considered a responder at the first time the response criteria of (confirmed) PR or CR are met. It is clinically equivalent if BOR is achieved early or late. The scientific question refers to a setting when achieving response is the primary treatment intention regardless of the time to onset. TTR provides the supportive information how quick onset of BOR occurs.

- **Confirmation of response requires following patients beyond the onset of initial response.** Per RECIST 1.1, it is recommended to confirm responses in single-arm studies. If confirmation is requested, a minimum follow-up of at least 1 scheduled assessment is required after a first response is achieved. The requirement to confirm an initial response is not intended to assess durability of response like DOR but to rule out potential measurement errors of the initial response assessment[3]. In early-stage clinical studies with limited follow-up, late onset of response increases the risk that no confirmation assessment can be performed and thus, patients are considered non-responders.

- **BOR does not take into account the disease dynamic**. Depending on the disease dynamic and the type of therapy a responder might improve from a PR to a CR with further therapy and may retain the response for a longer period. Therapies leading to transient responses with short duration result in the same ORR as a therapy which achieves durable responses. Therefore, DOR provides supportive information if we want to know if BOR is durable.

- **Not documenting at least partial response is considered failure**. Despite BOR is not a time-to-event endpoint it is based a related concept. Patients are typically followed for response assessments until criteria for PR or CR are met and further until progression or death is documented. Accordingly, intercurrent events like start of new antineoplastic therapies may prevent patients from being followed for further response assessments. Like for time-to-event endpoints, onset of response after initiating new anticancer therapies might be attributable to the effect of the new therapy and not to the study treatment. Thus, assessments after start of new anticancer therapies would not be considered to determine BOR. Intercurrent events limit the time period where patients have the chance to observe at least PR in the sense of following a *while-on-treatment* strategy for start of new therapies, progression and death. This implies that not achieving or not documenting response within this time period is considered as failure to treatment. As we discuss in Section 3, the chosen intercurrent event strategy for ORR has implications on the intercurrent event strategies for DOR and TTR estimands.

A typical ORR estimand in a single-arm design could be defined as follows:

- Scientific question: What is the expected ORR of population *P* treated with *T* regardless of treatment discontinuation or modifications based on the BOR as documented while being alive, not experiencing progression and not switching to a new anticancer therapy?
- Population: Patients with indication of interest as defined by selection criteria
- Treatment: Treatment *T* of interest
- Variable: BOR by investigator assessment
- Intercurrent event handling
    a. start of new anticancer therapy in addition to initially assigned therapy or stopping initially assigned therapy and switching to new anticancer therapy: *while not having started new anticancer therapy*
    b. disease progression: *while not having evidence of progression*
    c. death due to any reason: *while being alive*
    d. Treatment discontinuation in the absence of new anticancer therapy: *treatment policy*
- Summary measure: ORR

## 3.    Duration of respose and time to response

Since the durability of responses and the time to onset of response cannot be inferred from the ORR, DOR and TTR provide important information in addition to ORR. Typically, both DOR and TTR are part of the product characteristics if a drug receives market approval based on ORR. Guidelines from major Health Authorities recommend describing DOR in the context of single arm Phase II studies[2],[7]. TTR is not explicitly mentioned in these cancer endpoint guidelines but it is commonly presented in study reports as well as of in product characteristics.

### 3.1    Duration of response

Duration of response (DOR) is commonly defined as the time from onset of response to progression or death due to any reason, whichever occurs earlier[3]. Typically, DOR is assessed regardless of study treatment discontinuation or treatment modifications of any kind. In case no event (progression or death) is observed at the end of study, the last response assessment that indicates absence of progression is considered as the last time point when a patient still benefits from the therapy. Of note, even when patients do not maintain a response, the clock for DOR continues to run until death or criteria for progression are met (e.g., when a patient achieves only SD after a confirmed PR).

The high-level question of interest for DOR is "What is the DOR with treatment *T* in patients from population *P*?" However, from our experience this question is not posed as a standalone question but in conjunction with other scientific questions. DOR is part of a double question: "What is the ORR with treatment *T* in patients from population *P*? Among responders, what is then the DOR?". Indeed, DOR is conditional on being a responder. We refer to this definition of DOR as the conditional DOR (cDOR). It allows one to characterize the quality of responses to be non-transient, e.g., "Not only does the new treatment have an ORR of 40% but the responses are also long-lasting. The conditional DOR among responders was 7 months". On an individual level, a patient prefers a drug on which she/he responds. Between two drugs for which a patient responds, all other criteria being equal, she/he prefers the drug with longer cDOR. One way to extend this preference relation to the population level is as follows. For two drugs *D1* and *D2* with the same ORR, cDOR could be used to informally decide which is the better one. If the cDOR is longer on *D1* compared to *D2*, and all other criteria being equal, then drug *D1* is preferred over *D2* ($D1 \succ D2$). Meanwhile, for two drugs, *D3* with the same cDOR but with higher ORR compared to *D2* (D3 $\succ D2$), *D3* is preferred.

But if *D2* has higher ORR but shorter cDOR compared to *D4* then it is not straightforward to say which of the drugs is better or when two drugs are equivalent ($D2 \approx D4$). Nevertheless, in principle such preferences could still be expressed (Figure 2). Note that even if two drugs have the same ORR and cDOR in *P* this does not mean they have the same efficacy. It might be that one of the drugs provides a benefit in subpopulation *P1* only, and the other drug provides a benefit only in subpopulation *P2*. Then the efficacy for a patient treated with either will depend on the subpopulation the patient belongs to.

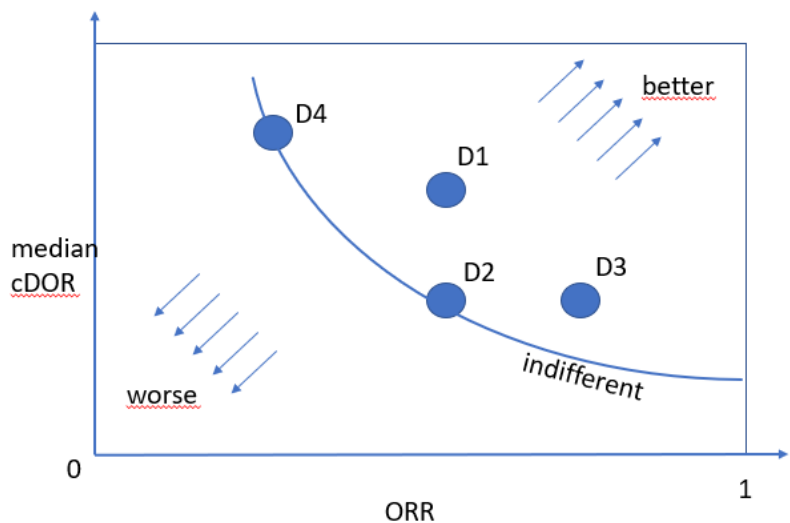

**Figure 2:** Preference relations of drug $D_i$ for ORR, median cDOR on population level

Considering cDOR as supportive information to ORR, we might not be interested to fully characterize the DOR distribution which requires sufficient follow-up. Instead of asking for the median cDOR we may want to know the proportion of responses with a duration reaching or exceeding a clinically meaningful time interval *i*. This approach informs a statement like "Not only does the new treatment have an ORR of 40% but the responses are also long-lasting. The majority of responders achieved PR or CR that lasted at least 6 months".

### 3.1.1 Traditional duration of response estimands

In a Phase I expansion cohort, we may be interested in an experimental treatment *T* for comparison to a fixed reference value of efficacy. In Phase II, we may be interested in accelerated approval for a new treatment *T* based on a comparison to a randomized control group *C*. The population *P* will be defined by selection criteria. For cDOR, the estimand focuses on responders only.

The variable that we measure at the patient level is the time from onset of first response (PR or CR) to progression or death due to any reason. Since we consider duration of response as supportive information to ORR, we may want to apply the same considerations for the variable definition as for ORR:

- If response requires a confirmation assessment, the clock of cDOR starts at the initial assessment of PR or CR that was subsequently confirmed
- Duration of response should be based on the same assessor as the corresponding response (investigator or IRC).

Relevant intercurrent events for cDOR that can prevent observing the endpoint or affect its interpretation may include the following:

- Treatment discontinuation or modification
- Start of new anticancer therapy

Progression and death are considered as events of interest and are handled using a composite strategy as part of the variable definition: Duration of response ends with progression or death, whatever occurs first. In conditions where life expectancy is short and the differentiation of the reason of death between study indication and other reasons is difficult, death due to any reason might be considered as event. If cure can be anticipated for a part of the study population, we might be interested to consider only death due to the study indication as an event.

In oncology, treatment discontinuation or modification is typically handled by a treatment policy strategy, i.e. we are interested in the duration of response regardless of treatment discontinuation or modification during or even prior to onset of response. If a response starts after the start of a new anticancer therapy, then it may be assumed that the response is attributable to the new therapy and not induced by the experimental therapy $T$ (see Section 2). If, during a response, $T$ would be changed with no documentation of progression (e.g. due to toxicity), then according to the same considerations, it may no longer be clear if the extent of cDOR is due to $T$ or the new anticancer therapy. The following considerations should be taken into account, assuming the start of new therapy occurred after onset of response and prior to progression or death:

- A **treatment policy** strategy where we are interested in the cDOR regardless of subsequent therapy may therefore not be appropriate for cDOR.
- If we are interested in cDOR if patients had not taken subsequent therapy, then we choose a **hypothetical strategy**.
- If we are interested in the cDOR while patients are on the experimental treatment $T$, then we choose a **while-on-treatment strategy**.
- If we are interested in a treatment failure interpretation, where cDOR ends with start of subsequent therapy, we choose a **composite strategy**. This modified cDOR would be defined as the time from response to start of subsequent therapy, progression or death.
- If we are interested in the DOR in a population that would not take subsequent therapy, then we would choose a **principal stratum** strategy.

One reason to prescribe new anticancer therapy is evidence of clinical progression. The treating physician may decide that the patient has progressed but the progression is not (yet) documented via objective criteria for progression, e.g. RECIST 1.1. Another reason for switching to a new anticancer therapy is toxicity of $T$. In principle we could distinguish between these cases and choose a different intercurrent event strategy for each case. Estimands for cDOR have not been discussed in the literature in detail so far. From published analyses we can infer the estimands that correspond to cDOR analyses. The typical cDOR estimand uses the following intercurrent event strategies (Table 1):

| **Intercurrent event** | **Strategy** |
|---|---|
| Progression, death | Composite |
| Treatment discontinuation or modification | Treatment policy |
| New anticancer therapy | Hypothetical |

**Table 1:** Typical intercurrent event handling for cDOR

There might be benefit from a disease-modifying therapy beyond end of treatment. So we would follow patients for response assessments beyond permanently stopping treatment and acknowledge that onset or duration of response is attributable to the treatment (*treatment policy* strategy). For the hypothetical approach, patients with new anticancer therapy are censored at the last adequate response assessment prior to the start of new therapy. cDOR events will not be observed if the study ends before we can observe them. Patients are then censored at the last adequate tumor assessment prior to the analysis cut-off date.

As summary measure, typically Kaplan-Meier estimates of the median cDOR as well as the cDOR rate at landmark timepoints of interest are provided (Table 2).

Defining a principal stratum strategy might not be the appropriate approach when we wish to focus on responders only to address cDOR. This strategy would require a reliable identification of patients who will respond post-baseline. A subgroup analysis of responders is generally not consistent with a principal stratum strategy. Therefore, we would rather refer to the population attribute of cDOR and cTTR to the subgroup of responders and not referring to a principal stratum strategy to address intercurrent events that prevent patients to be considered as responders.

### 3.1.2 Alternative duration of response estimands

We may be interested in the unconditional duration of response. Such an estimand addresses a question which we may have at the time when a treatment decision is to be made of what is the effect when we do not know if a patient will be a responder. Obviously, the population attribute of the corresponding estimand would refer to the whole study population. Intuitively we may want to integrate both ORR and the duration of response together. Huang[8] called this the *expected duration of response* (EDOR). The definition of the EDOR estimand is summarized in Table 2. For the estimation, a multi-state model is proposed to describe the *probability of being in response* at time *t*. For each time point *t* after an index date (e.g. start of treatment) patients have a certain probability to be not yet in response, be currently in response or be no longer in response after having progressed or died. This model brings together the time to response, duration of response and progression-free survival (PFS). It allows one to define the EDOR as the area under the probability of being in response function. The model provides the conditional overall response rate (cORR) for every time point *t* after the index date. Another approach assessing the whole population is suggested by the EMA (2017) guidance[7]. It is defined as the *time in response:* On patient level, DOR is set to 0 for a patient with no response, and equals the cDOR definition if the patient is a responder.

A problem is that we can't perform a valid statistical comparison of the cDOR between two treatments *T* and *C* by simply comparing unadjusted summary statistics for cDOR since the subpopulation of responders to treatment *T* may differ from the subpopulation of responders to treatment *C*. If one wishes to perform comparisons across treatment groups, unconditional estimands like EDOR or time in response are recommended as described above (Table 2).

Korn[9] addressed the concern of different population sizes when comparing cDOR across treatment groups. Specifically, they identified and removed responding patients in *T* who are least likely to respond had they received *C* (or non-responding patients in *C* who are most likely to respond had they received the *T*). This approach leads to treatment subsets of the same size as determined by the number of responders in *C*. However, it remains unclear if there is a meaningful scientific question that can be formulated corresponding to this approach.

The principal stratification framework offers another approach for a comparison of cDOR across treatment groups. The key question is to estimate the difference between duration of response in the group of patients that would have responded under both treatment groups. The efficiency of the approach depends on whether meaningful covariates can be collected to predict the probability of response in each treatment group[10].

|  | **Traditional cDOR (1)** | **cDOR: Alternative 1 for new anticancer therapy (2)** | **cDOR: Alternative 2 for new anticancer therapy (3)** | **Expected DOR (4)** | **Time in response (5)** |
|---|---|---|---|---|---|
| Scientific question | Among responders from population *P* treated with *T*, what is the time from onset of response to progression or death, regardless of treatment discontinuation and assuming absence of new anticancer therapy? | Among responders from population *P* treated with *T*, what is the time from onset of response to progression or death, regardless of treatment discontinuation and new anticancer therapy? | Among responders from population *P* treated with *T*, what is the time from onset of response to progression or death while not starting new anticancer therapy? | What is the expected time in response of population *P* treated with *T* over an observation period of x months regardless of treatment discontinuation and assuming absence of subsequent therapies? | What is the time in response of population P treated with T regardless of treatment discontinuation and assuming absence of subsequent therapies? |
| Treatment | Treatment *T* of interest (and comparator, if applicable) | like (1) | like (1) | like (1) | like (1) |
| Population | Patients with indication of interest as defined by selection criteria who achieved response | like (1) | like (1) | Patients as defined by selection criteria | Patients as defined by selection criteria |
| Variable | Time from onset of PR or CR until progression or death due to any reason, whichever is earlier | like (1) | like (1) | Time from onset of PR or CR until progression or death due to any reason, whichever is earlier (0 for non-responders) | Time from onset of PR or CR until progression or death due to any reason, whichever is earlier (0 for non-responders) |
| Intercurrent events | Treatment discontinuation: treatment policy New anticancer therapy: | Treatment discontinuation: treatment policy | Treatment discontinuation: treatment policy New anticancer therapy: *while not having* | Treatment discontinuation: treatment policy Subsequent therapy: *hypothetical strategy* | Treatment discontinuation: treatment policy New anticancer therapy: |

|  | Traditional cDOR (1) | cDOR: Alternative 1 for new anticancer therapy (2) | cDOR: Alternative 2 for new anticancer therapy (3) | Expected DOR (4) | Time in response (5) |
|---|---|---|---|---|---|
|  | *hypothetical strategy* | New anticancer therapy: *treatment policy* | *started new anticancer therapy* |  | *hypothetical strategy* |
| Population-level summary | Median | Median | Median | Expected DOR | Median |
| Example | Levy (2022)[11] | Shah (2021)[12]: sensitivity analysis |  | Robertson (2016)[13] | Herbst (2020)[14] |

**Table 2:** Duration of response estimands

P – population, T – treatment, PR – partial response, CR – complete response, DOR – duration of response, cDOR – conditional DOR

## 3.2 Time to response

In general, a drug *D1* leading faster to response is preferable over a drug *D2* provided that ORR and other criteria are identical. Like for cDOR, we may only be interested in the conditional TTR (cTTR) for responders and consider it as supportive information to characterize ORR. cTTR may be presented in product labels as part of a double question: „What is the ORR with treatment *T* in patients from population *P*? Among responders, what is then the TTR?". cTTR characterizes how fast response can be achieved next to durability, e.g., „The new treatment has an ORR of 77%. Responses are achieved quickly (median 2 cycles) and the responses are durable (median 13 months)".

On an individual level, a patient prefers a drug on which she/he achieves a response. Among two drugs on which a patient responds, she/he prefers the drug with longer cDOR. If the ORR and cDOR is the same for two drugs, we might then think that the drug with the earlier onset of response should be preferred.

### 3.2.1 Traditional estimand for time to response

The population *P* is defined by selection criteria and may consist of responders only. The endpoint is the time from start of therapy or randomization to onset of response. Confirmation requirements for response and the assessor (investigator or IRC) will follow that of the corresponding ORR estimand. If we are interested in confirmed responses then the cTTR would be the time from start of treatment until the initial PR or CR that is subsequently confirmed. For instance, the case presented in Figure 1 has an initial PR which was then confirmed by a subsequent CR. The summary measure is the median cTTR. Intercurrent event strategies for cTTR follow the considerations of the strategies which have been implemented for ORR. If we are interested in responses regardless of subsequent new therapies then this implies that we would also follow a treatment policy strategy for cTTR.

- Response is the event of interest for cTTR and thus, part of the variable definition. Under a **composite strategy**, additional events reflecting treatment success may be added (e.g., receiving a stem cell transplant in hematologic malignancies).
- A **treatment policy** strategy could be implemented when we are interested in the cTTR regardless of new anticancer therapy. We may argue that this approach may only be appropriate if we follow the treatment policy strategy also for ORR.
- If we are interested in the cTTR assuming absence of new anticancer therapy, then we choose a **hypothetical strategy**.
- If we are interested in the cTTR while patients are on *T* then we choose a **while-on-treatment strategy**.

### 3.2.2 Alternative time to response estimands

The attributes of a typical cTTR estimand are outlined in Table 3 (estimand 1). A hypothetical strategy for patients who died or progressed might be difficult to justify since progression and death will ultimately prevent from observing response and thus, we cannot assume same hazards for response as for patients who are alive without progression. In publications we may see the approach to set the censoring date to the maximum follow-up time of the study at the population level (Table 3, estimand 2). This approach rearranges the ordering of event and censoring times. However, while this reordering avoids technically that the median KM estimate is impacted by censoring times for progression or death, there is no clinically meaningful interpretation of using the maximum follow-up time of the trial. This approach also implies that we may follow the treatment policy strategy regarding terminal events. Such strategies are not recommended by the ICH E9 (R1) addendum. A *while not having evidence of progression* or *while being alive* strategy could be considered regarding progression or death when responses are only achievable prior to progression and death (Table 3, estimand 3). The whole population would be of interest regardless of whether response was achieved. We may define progression, death and start of antineoplastic therapies as competing events. For further details we refer to Rufibach (2019)[15]. Instead of the median time to response, we might be interested to understand if patients have achieved response within a certain time interval. The summary measure is then the proportion of patients with response in the interval of interest.

Progression, death and new therapies can be considered as alternative states in a multistate model[16] or as competing events for response and we would be able to observe response only while patients have not progressed, did not start a new anticancer therapy and are alive. Patients can either transition to response (prior to start of a new anticancer therapy, progression or death) or directly to new therapies, progression or death without achieving response. The transition probability to response can be estimated by the cumulative incidence function (CIF). This approach informs us about the probability of having achieved response within time *t*. We intend to estimate the difference between time to response in the group of patients that would have responded under both treatment groups.

In a controlled study, no valid statistical comparisons of the cTTR of *T* and *C* can be performed by comparing unadjusted summary statistics for cTTR since the subpopulation of responders may be different across treatment groups.

|  | **Traditional cTTR (1)** | **TTR: treatment policy (2)** | **TTR: while on treatment (3)#** |
|---|---|---|---|
| Scientific question | Among responders from population *P* treated with *T*, what is the time to response regardless of treatment discontinuation? | What is the time to response for population *P* treated with *T* regardless of progression, death and treatment discontinuation? | What is the time to response for population *P* treated with *T* while not experiencing progression or death, and not switching to new anticancer therapy, but regardless of treatment discontinuation? |
| Treatment | Treatment *T* of interest (and comparator, if applicable) | Treatment *T* of interest (and comparator, if applicable) | Treatment *T* of interest (and comparator, if applicable) |
| Population | Patients with indication of interest as defined by selection criteria who achieved response prior to progression, death and new anticancer therapy | Patients with indication of interest as defined by selection criteria | Patients with indication of interest as defined by selection criteria |
| Variable | Time from start of therapy to onset of PR or CR | Time from start of therapy to onset of PR or CR. TTR is set to maximum follow-up time in the trial in case of progression or death, or to last assessment otherwise | Time from start of therapy to onset of PR or CR |
| Intercurrent events | Treatment discontinuation: treatment policy | Treatment discontinuation: treatment policy Progression, death: ***treatment policy***[1] | Treatment discontinuation: treatment policy Progression: ***while not having evidence of progression*** Death: ***while being alive*** New anticancer therapy: ***while not having started new anticancer therapy*** |
| Population-level summary | Median | Median | Median |
| Example | Skoulidis (2021)[17] | | |

**Table 3:** Time to response estimands

P - population, T – Treatment, TTR – Time to response, cTTR – conditional TTR, PR – partial response, CR – complete response
[1] since no follow-up beyond terminal events is possible, the maximal follow-up time in the trial is sometimes used
# Terminology for the *while on treatment* strategy depends on the intercurrent event of interest as pointed out in ICH E9 (R1) addendum.

## 4. Case study in mantle cell lymphoma

We consider a single-arm phase 2 study in mantle cell lymphoma (MCL) as a case study. This study includes data from 30 patients. The primary endpoint was ORR, defined as partial (PR) or complete response (CR) as per the Revised IWG Criteria for non-Hodgkin's lymphoma. Tumor assessments were performed during screening and repeated at cycles 3, 5, and 7 and then every three cycles until disease progression or death. Every cycle lasted 28 days. No confirmation of response is required. The secondary endpoints included DOR and TTR. DOR was measured from the time when the criteria for PR or CR were met until progressive disease or death was documented and TTR was defined as the time from start of therapy until onset of response. The dataset does not include information on intercurrent events like start of new antineoplastic therapies. Thus, we assume that all patients were followed until progression or the analysis cut-off date with no start of new antineoplastic therapies. Administrative censoring was applied when progression or death was not documented at the analysis cut-off date. In total, n=23 patients responded to study treatment (ORR: 76.7%).

First, we assess the traditional conditional DOR estimand (Table 2, estimand 1). The median cDOR is not yet reached and thus, the majority of patients are still in response at the time of the analysis cut-off date. We may conclude that responses are durable with 86% (95% CI: 73%, 100%) patients still in response after 6 months as per KM estimates (Table 4, row 1). The corresponding KM plot is presented in Figure 3.

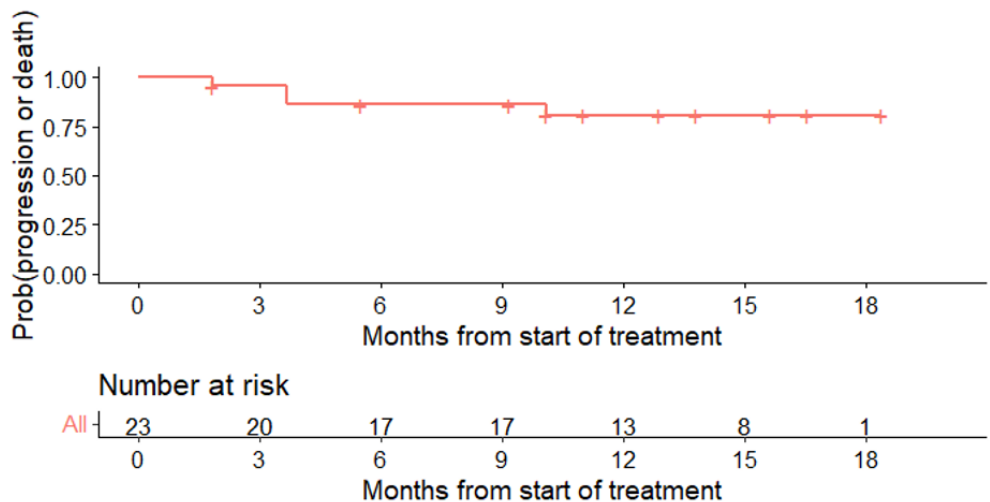

**Figure 3:** MCL case study - Kaplan-Meier plot for cDOR

We may be interested in the unconditional DOR based on all patients: We ask for the *time in response* (as defined by EMA 2017[7]) and set DOR to 0 months for patients without response (Table 2, estimand 5). Results are provided in Table 4, row 2. The similar question is assessed by the expected mean DOR (Table 4, row 4) for which we use a cut-point of 9 months (Table 2, estimand 4). As supportive information we are interested in the (unconditional) probability of being in response (PBIR) at 6 months (Table 4, row 3).

The traditional cTTR estimand (Table 3, estimand 1), the onset of response amongst responders occurred early after start of therapy (Table 5, row 1): The median conditional time to response (cTTR) occurred at the first scheduled response assessment after 1.84 months (2 cycles). The corresponding Kaplan-Meier plot is presented in Figure 4.

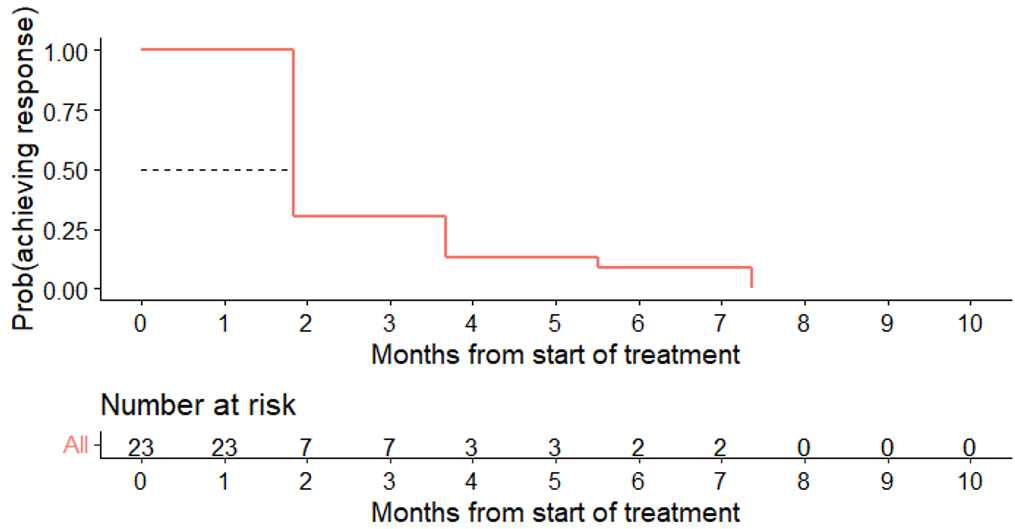

**Figure 4**: MCL case study - Kaplan-Meier plot for cTTR

Based on all patients, we censor at the maximum follow-up time of the study if patients experience progression prior response and at the last response assessment otherwise (Table 3, estimand 2; Table 5, rows 7 and 8). Since it is not possible to document response after progression, the results of the treatment policy strategy might not be interpretable. The 6-month time to response as derived from the cumulative incidence function is presented in Table 5, row 9, corresponding to estimand 3 (Table 3).

| Number | Description | Population | Result |
| --- | --- | --- | --- |
| 1 | cDOR: KM 6-month estimate (95% CI) | Responders | 0.86 (0.73,1.00) |
| 2 | DOR: KM 6-month estimate (95% CI) | All patients | 0.66 (0.51, 0.86) |
| 3 | Probability of being in response at Month 6 | All patients | 0.63 (0.48, 0.75) |
| 4 | Mean expected duration of response [months] | All patients | 4.64* |

**Table 4**: Results of various duration of response estimands in the MCL case study

DOR – Duration of response, cDOR – conditional duration of response, EFS – event-free survival, nr – not reached; *truncated 9.2 months

| Number | Description | Population | Result |
| --- | --- | --- | --- |
| 5 | cTTR: KM estimate [months] (median, 95% CI) | Responders | 1.84 (1.84, 3.68) |
| 6 | cTTR: KM 6-month estimate (95% CI) | Responders | 0.09 (0.02, 0.33) |
| 7 | TTR: Median TP [months] (95% CI) | All patients | 1.84 (1.84, 7.36) |
| 8 | TTR: KM 6-month TP estimate (95% CI) | All patients | 0.27 (0.15, 0.50) |
| 9 | TTR: 1-CIF 6-months estimate (95% CI) | All patients | 0.27 (0.10, 0.44) |

**Table 5**: Results of various time to response estimands in the MCL case study

TTR – time to response, cTTR – conditional time response, CIF – cumulative incidence function, TP – TTR is set to maximum follow-up for patients with progression prior to response and to the last response assessment otherwise

A swimmer plot presents the individual follow-up from start of therapy by best overall response (Figure 5). It illustrates patients who are still in response at the time of the analysis cut-off date using an arrow symbol or events of interest occurred during the course of the study.

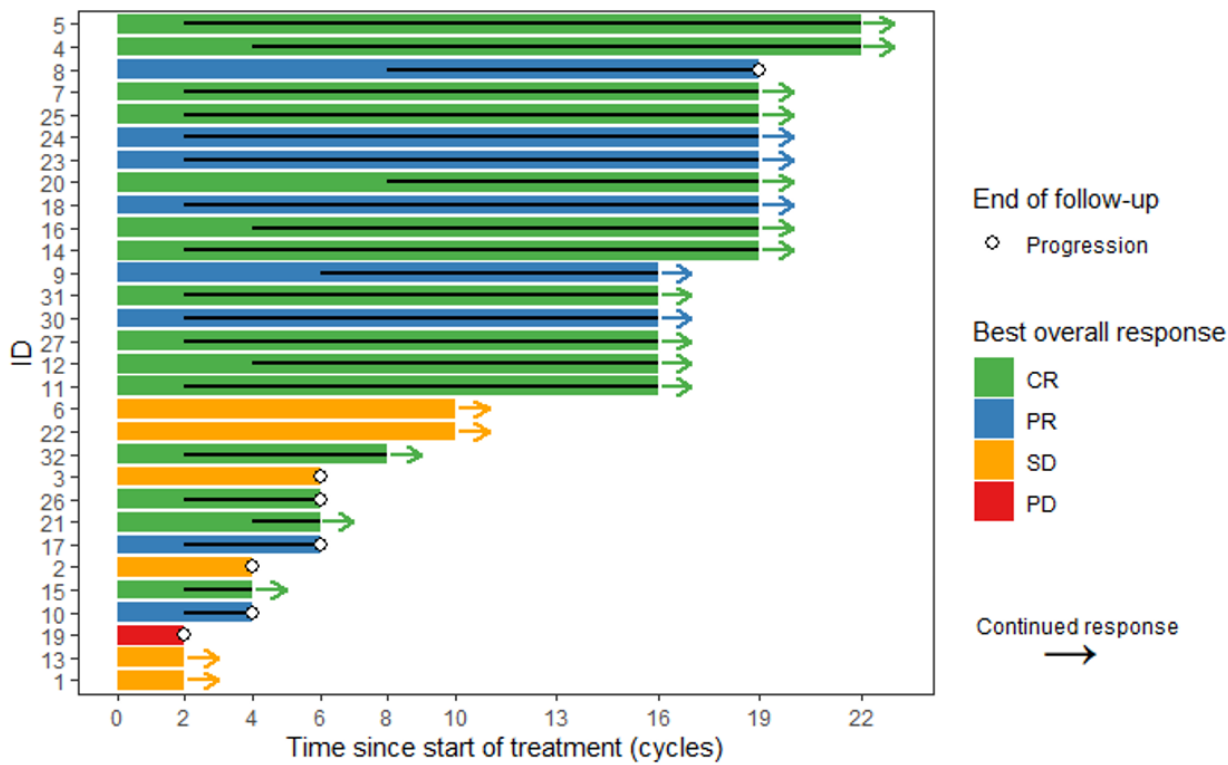

**Figure 5**: Swimmer plot illustrates efficacy follow-up, time to and duration of response (MCL case study)

# 5. Discussion and recommendations

Duration of response and time to response are common secondary efficacy endpoints in early-stage clinical trials in oncology when ORR is the primary endpoint. The corresponding results are reflected in product labels in case of accelerated approvals. However, per a non-systematic PUBMED review of clinical trial publications in oncology, there is in general no clear definition provided about the DOR and TTR estimands. As summarized in Section 3 and illustrated by the MCL case study, different estimands of DOR and TTR address different scientific questions. To ensure appropriate interpretation we recommend to define in study protocols and statistical analysis plans at least the scientific question of interest, the population, the variable attributes and the strategy considered to deal with intercurrent events.

Rufibach (2019)[15] points out that common endpoints in oncology like PFS, OS, EFS (event-free survival), DFS (disease-free survival), and DOR "are basically received by varying handling of clinical events for the variable and intervention effect, as well as the population attribute for DFS and DOR". Also TTR fits into this concept.

To characterize durability of response then, we might be interested in BORs with a minimum duration of *t*. In fact, this endpoint is a generalization of the confirmed response endpoint, where instead of a single confirmation of a response after for example, 6 weeks, we request typically a longer duration of response. The length of such an interval should be clinically justified.

In publications of clinical trials, it is common to assess both DOR and TTR for responders only. This practice is also adopted by Health Authorities for product labels based on single-arm studies. We reconstructed the corresponding "traditional" cDOR and cTTR estimands (Table 2, estimand 1 and Table 3, estimand 1) and recommend to follow this practice in the clinical context of such single-arm phase 2 studies with a primary ORR estimand. These estimands further characterize the primary ORR estimand. Since cDOR and cTTR are supportive estimands, sensitivity and supplementary analyses might not be required.

Well-defined DOR and TTR estimands are a prerequisite for proper data collection. The visit schedule should be designed in a way that information can be collected and harmonized with relevant response assessment intervals. In particular, when response assessments require invasive maneuvers like bone marrow aspirates or imaging, trialist need to find a balance between a regular assessment and minimizing discomfort and potential harm to patients. Unscheduled response assessments might be allowed to detect deterioration or improvements based on clinical features. Moreover, depending on the estimand definition, follow-up for cDOR might be required after discontinuation of study treatment. As DOR addresses durability of responses, the data collection schedule should ensure to follow patients for a clinically relevant time interval after the last patient has started therapy. Since in early phase studies there is often limited time to follow patients, the summary measure may address a KM estimate for the clinically meaningful time interval instead of providing the median KM.

In general, DOR is assessed in conjunction with a primary ORR estimand. There are clinical assumptions implied in the typical ORR estimand which are also relevant for cDOR and cTTR. We therefore recommend to make these assumptions explicit and assess their relevance in the light of the specific clinical setting. Considerations for intercurrent event strategies of cDOR are typically guided by those that are relevant to ORR. For instance, if the ORR estimand addresses start of new anticancer therapy by treatment policy, the same clinical considerations might apply for dealing with new antineoplastic therapies for cDOR and cTTR. We would advise teams to discuss and to develop intercurrent event strategies together for ORR, cDOR and cTTR.

Non-adherence to response assessments requires further consideration. If non-adherence does not relate to intercurrent events we propose considering these cases as missing data. A hypothetical strategy could be followed in case of more than 1 missing assessment prior to the event of interest as suggested for a similar situation regarding progression-free survival[2]. However, this might be difficult to justify as we know that these patients experience the event of interest. Therefore, we recommend considering the event regardless of preceding missing assessments for both cDOR and cTTR.

Interpretation of cDOR and cTTR together with ORR has limitations. From a clinical perspective, one could argue that a therapy *A* resulting in a higher ORR than therapy *B* should be preferred over a therapy *B* with lower ORR but longer cDOR as discussed in Section 3.1. However, patients' preference might depend on additional considerations. Patients might prefer therapy *B* with a lower ORR but longer responses compared to a therapy *A* with a higher chance of response but shorter duration. Thus, there might be no general applicable compromise between ORR and cDOR (Figure 2). Shafrin (2017)[18] investigated patients' and physicians' preferences with regard to a therapy *X* resulting in a fixed overall survival benefit or a therapy *Y* with a variable overall survival that could exceed that of the therapy *X* – but could also be less. While physicians preferred the fixed overall survival therapy *X*, patients expressed preference to therapy *Y*. So patients may be willing to take a higher risk to obtain a better outcome when comparing treatment options. Similar conflicting preferences might be relevant when judging therapies with different outcomes of ORR vs cDOR. This justifies to use the traditional cDOR and cTTR estimands primarily in single-arm studies.

Despite early onset of response could be beneficial for patient's quality of life, there is no general evidence that cTTR is a predictor of overall survival. In some indications early onset of response is a relevant treatment objective (e.g., in acute myeloid leukemia[19]). However, in other indications there is no clear relationship. In contrast, a short cTTR might even result in a shorter cDOR which does not seem to be a reasonable preference. Therefore, the assessment of cTTR needs to be done in the context of other relevant characteristics covering ORR and time-to-event assessments.

The cDOR and PFS estimand definitions are related, in particular if responses occur early after starting therapy. However, for cDOR we are interested only in the subset of responders. Unlike for PFS, no statement about a treatment benefit is made about non-responders. Moreover, patients might still benefit from a therapy which delays progression or death even when no response was achieved. Despite the definition has similarities, the underlying scientific questions of cDOR and PFS are different. The relevance of cDOR might be specific as supplemental information regarding the durability of the ORR.

As highlighted by McCaw (2020)[20], comparisons of cDOR across treatment groups are not appropriate. We cannot estimate a causal treatment effect by comparing the difference of cDOR across treatment arms as the population of responders is not comparable. Related questions can be addressed by randomized, controlled studies with acknowledged time to event estimands like PFS or disease-free survival (DFS). In case of phase 2 studies with 2 groups with limited follow-up we recommend assessing the expected duration of response (EDOR, Table 2). This estimand addresses integrates both the response and its duration. Hu (2021)[21] explored EDOR in randomized phase 2 trials and found that it is potentially more sensitive than PFS and ORR in estimating the subsequent phase 3 study outcomes. As an example of a published EDOR estimand is the FALCON trial[13].

Considering that cDOR and cTTR primarily characterize a primary ORR estimand, there is in general no need for more specific sensitivity and supplementary analyses in a typical single-arm trial. However, such analyses could be motivated by sensitivity or supportive analyses for ORR. This applies when different intercurrent events strategies are applied and we are interested to assess DOR and TTR with a consistent approach. Potential supplementary or sensitivity analyses may have impact on data collection procedures: For instance, if we are interested in assessing DOR regardless of starting new therapies in a supplementary analysis, then continuing follow-up for efficacy assessments beyond start of new therapies should be considered.

Graphical tools can be helpful to summarize and communicate DOR and TTR outcomes. In smaller phase 2 studies, swimmer plots visualize not only the onset and duration of responses but also study treatment termination and modifications as well as the post-treatment efficacy follow-up status (Figure 5). The onset of intercurrent events can also be shown. The onset response can be displayed as well as the evolution of responses along the study, e.g., if patients improve from an early PR to a CR. For cDOR and cTTR we also recommend to present the KM curves to show the whole distribution which is more informative than just providing the median KM estimate.

# 6.    Code Availability Statement

Code that implements all the methods discussed in this paper is available at
https://oncoestimand.github.io/dor/dor.html

# 7.    Data Availability Statement

No new clinical data is presented in this manuscript. The case study data is available via the code (Section 6).

# 8.    Acknowledgements

This paper has been written within the industry working group estimands in oncology, which is both, a European special interest group "Estimands in oncology", sponsored by PSI and European Federation of Statisticians in the Pharmaceutical Industry (EFSPI) and a scientific working group of the biopharmaceutical section of the American Statistical Association. Details are available on www.oncoestimand.org. We thank David Wright for helpful comments.